\documentclass[12pt,a4paper]{article} %
\usepackage{feynman} 
\usepackage{graphicx} %

\begin{document}
\begin{center}
{\large \bf Critical behavior of disordered systems with replica symmetry breaking}
\vspace{0.5cm}

{\bf P. V. Prudnikov, V. V. Prudnikov \\
Omsk State University, \\
Omsk, 644077 Russia}
\end{center}

\begin{abstract}
A field-theoretic description of the critical behavior of weakly disordered systems with
a $p$-component order parameter is given. For systems of an arbitrary dimension in the
range from three to four, a renormalization group analysis of the effective replica
Hamiltonian of the model with an interaction potential without replica symmetry is given
in the two-loop approximation. For the case of the one-step replica symmetry breaking,
fixed points of the renormalization group equations are found using the Pade-Borel
summing technique. For every value $p$, the threshold dimensions of the system that
separate the regions of different types of the critical behavior are found by analyzing
those fixed points. Specific features of the critical behavior determined by the replica
symmetry breaking are described. The results are compared with those obtained by the
$\varepsilon$-expansion and the scope of the method applicability is determined.
\end{abstract}

\section{Introduction}

When the renormalization group approach is
applied to describe the critical behavior of disordered
systems with quenched disorder, the method of replicas
\cite{1,2,3} is used to restore the translation symmetry of the
effective Hamiltonian describing the interaction of fluctuations.
However, in the studies \cite{4,5,6}, it was conjectured
that the replica symmetry could be broken in systems
with quenched disorder. In \cite{4,5,6}, the physical
concept of the occurrence of numerous local energy
minima in disordered systems with random transition
temperature was used to give a renormalization group
description of the $\phi^4$ model with the interaction potential
characterized by the broken replica symmetry. For
this purpose, the $\epsilon$-expansion technique was used in the
lower order of the theory. For systems in which the
number of components of the order parameter $p$ is less
than four, it was discovered that the breaking of replica
symmetry is the crucial factor in the critical behavior. It
was shown that, for $p$ in the range from one through
four, two modes of the system behavior are possible.
The first one determines a nonuniversal critical behavior,
which depends on seed values of the model parameters
and, ultimately, on the concentration of impurities
in the system. The second mode is characterized by the
absence of a stable critical behavior, as also is the most
interesting case of Ising ($p=1$) systems. Even though
the implications of these studies are very interesting,
the results of a field-theoretic description of certain
homogeneous and disordered systems in the two-loop
and higher order approximations based on the asymptotic
series summation techniques \cite{7} showed that the
stability analysis of various types of critical behavior
that uses the first-order terms of the
$\epsilon$-expansion can be
considered only as a coarse estimate, especially for
multivertex statistical models \cite{8}. For this reason, the
results of investigation of the replica symmetry breaking
(RSB) effects obtained in \cite{4,5,6} require revaluation
from the viewpoint of a more accurate approach.

To this end, we proposed in \cite{9,10}, in the framework of the field-theoretic
approach, a renormalization group description of the model of weakly disordered three-
and two-dimensional systems with the fourth-order interaction potential with respect to
the order parameter fluctuations, which determines the replica symmetry breaking. An
analysis of solutions to the renormalization group equations carried out in the two-loop
approximation with the sequential application of the summation technique for Pade-Borel
series showed that the critical behavior of three- and two-dimensional systems is stable
with respect to the relative influence of the RSB effects, and the former scenario of the
quenched disorder influence on the critical behavior is realized \cite{11}.

However, the scope of the results obtained in \cite{4,5}
remains unclear. In particular, it is interesting to establish
the threshold dimensions of the disordered system,
$d_c(p)$, that separate the domain of influence of RSB
effects from the critical behavior domains in which
these effects are insignificant. It is also interesting to
apply the renormalization group approach to analyze
the behavior of systems with replica symmetry breaking
in which no stable critical behavior exists and the
strong coupling mode occurs (see \cite{4,5}). A theoretical
analysis of this phenomenon is especially important
from the viewpoint of the possible manifestation of
RSB effects in strongly disordered systems and their
observation in computer models of the critical behavior
under the impurity concentration exceeding the impurity
percolation threshold when extended impurity
structures are formed in the system \cite{12}.

This paper is devoted to the consideration of the
above-mentioned problems. For weakly disordered
systems of an arbitrary dimension in the range from
three to four, an analysis of the critical behavior of the
model with an RSB potential is carried out based on the
renormalization group approach in the two-loop
approximation with the use of summation techniques.
Our analysis does not rely on the $\varepsilon$-expansion technique.

\section{Definition of the model
and the calculation procedure}

The model Ginsburg-Landau Hamiltonian, which
describes the behavior of a $p$-component spin system
with weakly quenched disorder near the critical point
has the form
\begin{eqnarray}
\label{H}
 H=\int d^dx\{\frac{1}{2}\sum_{i=1}^{p}[\nabla{\phi}_{i}(x)]^{2}+
\frac{1}{2}[\tau-\delta\tau(x)]\sum_{i=1}^{p}{\phi}_{i}^{2}(x)+
\frac{1}{4}g\sum_{i,j=1}^{p}{\phi}_{i}^{2}(x){\phi}_{j}^{2}(x)\}
\end{eqnarray}
where the random phase transition temperature has the
Gaussian distribution
$\delta\tau(x)$ with the variance $<<(\delta\tau(x))^{2}>>\sim u$,
which is determined by a positive constant $u$
and is proportional to the concentration of the
structure defects. The application of the conventional
replica method (see, for example, \cite{6}) makes it possible
to average over the temperature fluctuations $\delta\tau(x)$
and reduce the problem of the statistical description of a
weakly disordered system to the problem of the statistical
description of a homogeneous system with the
effective Hamiltonian
\begin{eqnarray}
\label{Hrepl}
 H_n=\int d^dx\{\frac{1}{2}\sum_{i=1}^{p}\sum_{a=1}^{n}[\nabla{\phi}_{i}^{a}(x)]^{2}+
\frac{1}{2}\tau\sum_{i=1}^{p}\sum_{a=1}^{n}[{\phi}_{i}^{a}(x)]^{2}+
\frac{1}{4}\sum_{i,j=1}^{p}\sum_{a,b=1}^{n}g_{ab}[{\phi}_{i}^{a}(x)]^{2}[{\phi}_{j}^{b}(x)]^{2}\},
\end{eqnarray}
Here, the index $a$ enumerates replicas (images) of the original homogeneous component in Hamiltonian
(\ref{H}); and the additional vertex $u$, which occurs in the interaction
matrix $g_{ab}=g\delta_{ab}-u$, , specifies the effective interaction
of fluctuations of the ($n\times{p}$) -component order
parameter through ground state of the system with the
configuration $\phi(x)=0$ ( at $T\geq T_c$), is performed at the
scale of the correlation length, which turns to infinity
the defect field. This statistical model is thermodynamically
equivalent to the original disordered model in the
limit  $n\rightarrow 0$.

The subsequent renormalization group procedure, which statistically takes into account the
contribution of long-wave fluctuations of the order
parameter relative to the at the transition temperature
$T_c$. This procedure makes it possible to analyze possible
types of the critical behavior and conditions of their
realization and calculate the critical indexes.

However, it was shown in \cite{4,5,6} that a macroscopically
large number of spatial regions with $\phi(x) \neq 0$
appears in the system due to fluctuations of the random
transition temperature at $[\tau - \delta\tau(x)]<0$. These regions
are separated from the ground state by potential barriers.
To describe the statistical properties of systems
with multiple local energy minima, the replica symmetry
breaking formalism (suggested by Parisi) was used
in \cite{4,5,6} by analogy with spin glasses \cite{9}. According to
the reasoning presented in \cite{4,5,6}, the statistical calculation
of the contribution of nonperturbation degrees of
freedom associated with the order parameter fluctuations
relative to the configurations of the field $\phi(x)$ at the
local energy minima results (when the replica procedure
is applied for the weak disorder) in the appearance
of additional interactions of the form in $\sum_{a,b}g_{ab}{\phi}_{a}^{2}{\phi}_{b}^{2}$
the effective replica Hamiltonian. Here, the final matrix
$g_{ab}$ is no longer replica symmetric with $g_{ab}=g\delta_{ab}-u$, but
rather has the RSB Parisi replica structure \cite{13}.
According to \cite{4,5,6,13}, the matrix $g_{ab}$ with the RSB
structure is parameterized (in the limit $n\rightarrow 0$ ) in
terms of its diagonal elements $\tilde{g}$ and the off-diagonal
function $g(x)$, which is defined on the interval $0 < x < 1$:
$g_{ab}\rightarrow (\tilde{g},g(x))$. Here, operations with the matrices
$g_{ab}$ are defined by the rules

\begin{eqnarray}
\label{mrule}
g_{ab}^{k}\rightarrow (\tilde{g}^{k};g^{k}(x)), \  \
(\hat{g}^{2})_{ab}=\sum_{c=1}^{n}g_{ac}g_{cb}\rightarrow (\tilde{c};c(x)), \  \
(\hat{g}^{3})_{ab}=\sum_{c,d=1}^{n}g_{ac}g_{cd}g_{db}\rightarrow (\tilde{d};d(x)),
\end{eqnarray}
where
\begin{eqnarray}
\label{mrule2}
\tilde{c}&=& \tilde{g}^{2} - \int_{0}^{1}dx g^{2}(x), \  \
c(x) = 2 [\tilde{g} - \int_{0}^{1}dy g(y)]g(x) - \int_{0}^{x}dy [g(x) - g(y)]^2, \\ \nonumber
\tilde{d}&=& \tilde{c}\tilde{g} - \int_{0}^{1}dx c(x)g(x), \  \
d(x) = [\tilde{g} - \int_{0}^{1}dy g(y)]c(x) +  [\tilde{c} - \int_{0}^{1}dy c(y)]g(x) -\\
&-& \int_{0}^{x}dy [g(x) - g(y)][c(x) - c(y)]. \nonumber
\end{eqnarray}
The replica symmetric situation corresponds to $g(x) = {\rm const}$,
which is independent of $x$.

\begin{figure}[p]
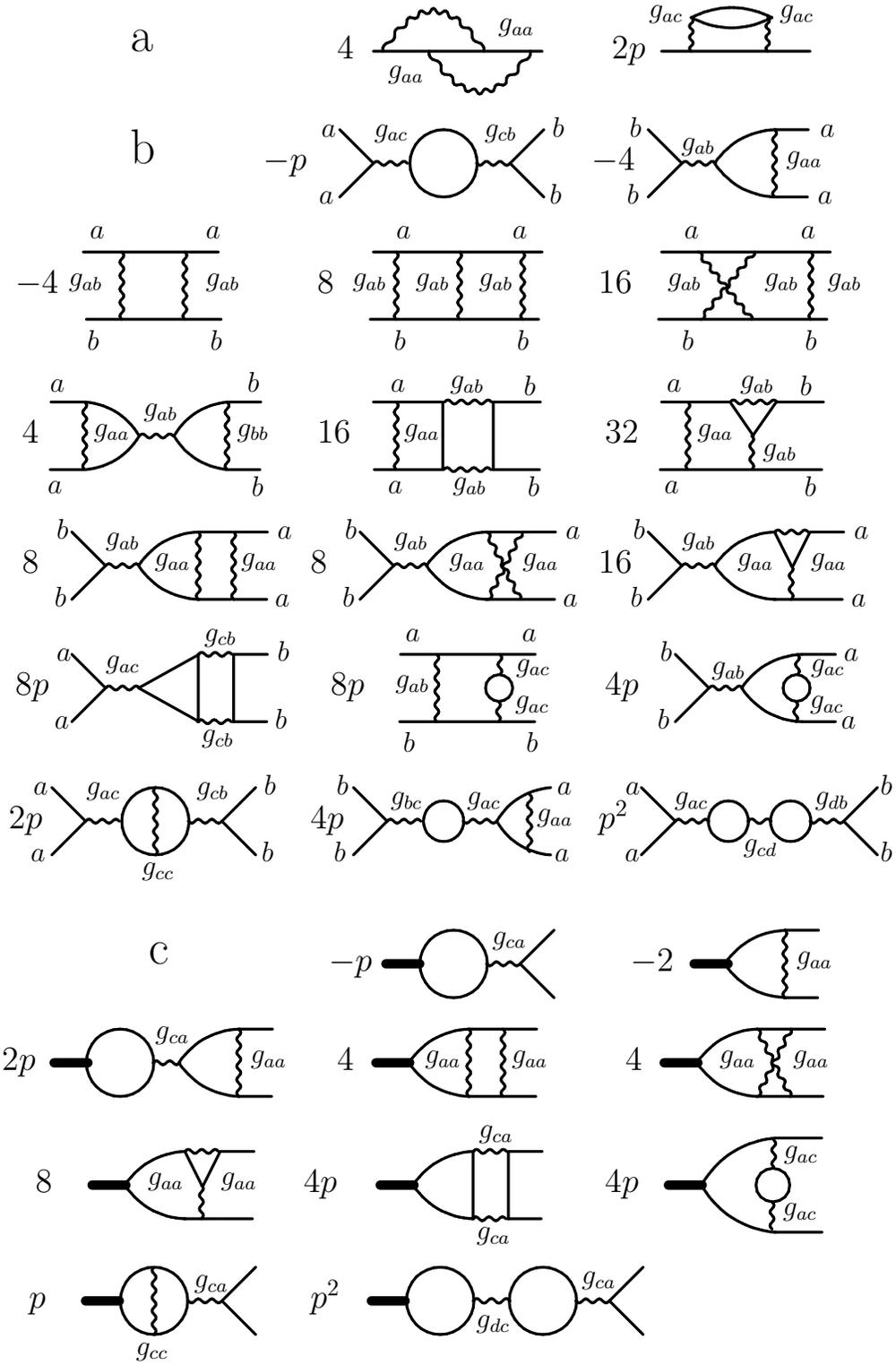

\Linewidth{1.2pt}
\Lengthunit=1cm
\Nhalfperiods=7
\vbox{
\vbox{

\hbox{          
\GRAPH(hsize=1) 
{}

\GRAPH(hsize=1) 
{}

\GRAPH(hsize=3) 
{
\mov(1.2,0.6){\LARGE a}
}
\GRAPH(hsize=1) 
{}

\GRAPH(hsize=3) 
{
\mov(0,0.5){\large$4$}
\mov(0.4,0.6){\lin(2.5,0)}
\mov(0.4,0.6){\wavearcto(1.5,0)[+1.6]}
\mov(0.95,0.6){\wavearcto(1.5,0)[-1.6]}
\mov(0.2,0.2){$g_{aa}$}
\mov(1.7,0.9){$g_{aa}$}
}
\GRAPH(hsize=1) 
{}

\GRAPH(hsize=3) 
{
\mov(-0.2,0.5){\large$2p$}
\mov(0.4,0.6){\lin(2.2,0)}
\Nhalfperiods=12
\mov(0.7,1.1){\arcto(1.2,0)[+0.5]\arcto(1.2,0)[-0.5]\wavelin(0,-0.5)}
\mov(1.7,1.1){\wavelin(0,-0.5)}
\mov(-0.2,1.1){$g_{ac}$}
\mov(1.6,1.1){$g_{ac}$}
\Nhalfperiods=7
}
} 
\vspace{0.3cm}
}
\Lengthunit=1cm
\Nhalfperiods=7
\vbox{

\hbox{          
\GRAPH(hsize=1) 
{}

\GRAPH(hsize=1) 
{}

\GRAPH(hsize=3) 
{
\mov(1.2,0.6){\LARGE b}
}
\GRAPH(hsize=1) 
{}

\GRAPH(hsize=3) 
{
\Nhalfperiods=10
\mov(-1.1,0.5){\large$-p$}
\mov(0.4,0.6){\wavelin(0.5,0)\lin(-0.5,0.5)\lin(-0.5,-0.5)}
\mov(1.3,0.6){\Circle(1)}
\mov(2.15,0.6){\wavelin(-0.5,0)\lin(0.5,0.5)\lin(0.5,-0.5)}
\mov(0,1){$g_{ac}$}
\mov(1.5,1){$g_{cb}$}
\mov(-1.1,0){$a$}
\mov(-1.2,1){$a$}
\mov(2.05,0){$b$}
\mov(1.95,1){$b$}
\Nhalfperiods=7
}
\GRAPH(hsize=1) 
{}

\GRAPH(hsize=3) 
{
\mov(-0.5,0.5){\large$-4$}
\Nhalfperiods=10
\mov(0.7,0.6){\wavelin(0.5,0)\lin(-0.5,0.5)\lin(-0.5,-0.5)}
\mov(1.05,0.6){\arcto(0.9,0.5)[+0.5]\arcto(0.9,-0.5)[-0.5]}
\mov(1.8,1.1){\wavelin(0,-1)\lin(0.5,0)}
\mov(1.65,0.1){\lin(0.5,0)}
\mov(1.7,0.6){$g_{aa}$}
\mov(-0.8,0){$b$}
\mov(-0.9,1){$b$}
\mov(1.75,0){$a$}
\mov(1.65,1){$a$}
\mov(-0.55,0.8){$g_{ab}$}
\Nhalfperiods=7
}
} 
\vspace{0.3cm}

\hbox{          

\GRAPH(hsize=1){}


\GRAPH(hsize=1) 
{}

\GRAPH(hsize=3) 
{
\Nhalfperiods=10
\mov(-0.5,0.5){\large$-4$}
\mov(0.37,1.1){\lin(2.0,0)}
\mov(0.8,1.1){\wavelin(0,-1)}
\mov(1.6,1.1){\wavelin(0,-1)}
\mov(0,0.1){\lin(2.0,0)}
\mov(-0.4,0.6){$g_{ab}$}
\mov(1.5,0.6){$g_{ab}$}
\mov(-0.4,-0.35){$b$}
\mov(-0.5,1.3){$a$}
\mov(1.15,-0.35){$b$}
\mov(0.95,1.3){$a$}
}
\GRAPH(hsize=1) 
{}

\GRAPH(hsize=3) 
{
\Nhalfperiods=10
\mov(-0.3,0.5){\large$8$}
\mov(0.4,1.1){\lin(2.5,0)}
\mov(0.6,1.1){\wavelin(0,-1)}
\mov(1.4,1.1){\wavelin(0,-1)}
\mov(2.2,1.1){\wavelin(0,-1)}
\mov(-0.2,0.1){\lin(2.5,0)}
\mov(-0.6,0.6){$g_{ab}$}
\mov(0.2,0.6){$g_{ab}$}
\mov(1.0,0.6){$g_{ab}$}
\mov(-0.4,-0.35){$b$}
\mov(-0.5,1.3){$a$}
\mov(1.15,-0.35){$b$}
\mov(0.95,1.3){$a$}
}
\GRAPH(hsize=1) 
{}

\GRAPH(hsize=3) 
{
\Nhalfperiods=10
\mov(-0.4,0.5){\large$16$}
\mov(0.4,1.1){\lin(2.5,0)}
\mov(0.8,1.1){\wavelin(0.8,-1)}
\mov(1.5,1.1){\wavelin(-0.8,-1)}
\mov(2.2,1.1){\wavelin(0,-1)}
\mov(-0.2,0.1){\lin(2.5,0)}
\mov(-0.2,0.6){$g_{ab}$}
\mov(1.1,0.6){$g_{ab}$}
\mov(1.9,0.6){$g_{ab}$}
\mov(-0.4,-0.35){$b$}
\mov(-0.5,1.3){$a$}
\mov(1.15,-0.35){$b$}
\mov(0.95,1.3){$a$}
}
} 
\vspace{0.3cm}

\hbox{          

\GRAPH(hsize=1){}


\GRAPH(hsize=1) 
{}

\GRAPH(hsize=3) 
{
\Nhalfperiods=10
\mov(-0.4,0.5){\large$4$}
\mov(1.2,0.6){\arcto(-0.8,0.5)[-0.5]\arcto(-0.8,-0.5)[+0.5]\wavelin(0.5,0)}
\mov(0.25,1.1){\lin(-0.5,0)\wavelin(0,-1)}
\mov(0.1,0.1){\lin(-0.5,0)}
\mov(1.3,0.6){\arcto(0.8,0.5)[+0.5]\arcto(0.8,-0.5)[-0.5]}
\mov(1.95,1.1){\lin(0.5,0)\wavelin(0,-1)}
\mov(1.8,0.1){\lin(0.5,0)}
\mov(-0.3,0.6){$g_{aa}$}
\mov(0.3,0.9){$g_{ab}$}
\mov(1.55,0.6){$g_{bb}$}
\mov(-1.4,-0.25){$a$}
\mov(-1.5,1.25){$a$}
\mov(1.35,-0.3){$b$}
\mov(1.15,1.25){$b$}
}
\GRAPH(hsize=1) 
{}

\GRAPH(hsize=3) 
{
\Nhalfperiods=10
\mov(-0.3,0.5){\large$16$}
\mov(0.4,1.1){\lin(1,0)}
\mov(0.6,1.1){\wavelin(0,-1)}
\mov(1.15,1.1){\lin(0,-1)}
\mov(0,0.1){\lin(1,0)}
\mov(0.9,1.1){\wavelin(0.7,0)}
\mov(0.75,0.1){\wavelin(0.7,0)}
\mov(1.35,1.1){\lin(0,-1)\lin(0.7,0)}
\mov(1.2,0.1){\lin(0.7,0)}
\mov(-0.25,0.6){$g_{aa}$}
\mov(0.3,1.3){$g_{ab}$}
\mov(0.2,-0.2){$g_{ab}$}
\mov(-0.9,-0.25){$a$}
\mov(-1.0,1.25){$a$}
\mov(0.9,-0.3){$b$}
\mov(0.7,1.2){$b$}
}
\GRAPH(hsize=1) 
{}

\GRAPH(hsize=3) 
{
\Nhalfperiods=10
\mov(-0.3,0.5){\large$32$}
\mov(0.4,1.1){\lin(1,0)}
\mov(0.65,1.1){\wavelin(0,-1)}
\mov(0.1,0.1){\lin(2.4,0)}
\mov(1.0,1.1){\wavelin(0.7,0)\lin(0.35,-0.5)}
\mov(1.55,1.1){\lin(-0.35,-0.5)\lin(0.7,0)}
\mov(1.05,0.6){\wavelin(0,-0.5)}
\mov(0.1,0.6){$g_{aa}$}
\mov(0.6,1.3){$g_{ab}$}
\mov(0.8,0.3){$g_{ab}$}
\mov(-0.8,-0.25){$a$}
\mov(-0.9,1.25){$a$}
\mov(1.0,-0.3){$b$}
\mov(0.8,1.2){$b$}
}
} 
\vspace{0.3cm}

\hbox{          

\GRAPH(hsize=1){}


\GRAPH(hsize=1) 
{}

\GRAPH(hsize=3) 
{
\Nhalfperiods=10
\mov(-0.4,0.5){\large$8$}
\mov(0.7,0.6){\wavelin(0.5,0)\lin(-0.5,0.5)\lin(-0.5,-0.5)}
\mov(1.05,0.6){\arcto(0.9,0.5)[+0.5]\arcto(0.9,-0.5)[-0.5]}
\mov(1.8,1.1){\lin(0.5,0)\wavelin(0,-1)}
\mov(1.65,0.1){\lin(0.5,0)}
\mov(2.05,1.1){\lin(0.5,0)\wavelin(0,-1)}
\mov(1.9,0.1){\lin(0.5,0)}
\mov(0.6,0.6){$g_{aa}$}
\mov(1.75,0.6){$g_{aa}$}
\mov(-0.4,0.9){$g_{ab}$}
\mov(-1.3,0){$b$}
\mov(-1.4,1){$b$}
\mov(1.7,0){$a$}
\mov(1.6,1){$a$}
}
\GRAPH(hsize=1) 
{}

\GRAPH(hsize=3) 
{
\Nhalfperiods=10
\mov(-0.4,0.5){\large$8$}
\mov(0.7,0.6){\wavelin(0.5,0)\lin(-0.5,0.5)\lin(-0.5,-0.5)}
\mov(1.05,0.6){\arcto(0.9,0.5)[+0.5]\arcto(0.9,-0.5)[-0.5]}
\mov(1.8,1.1){\lin(0.5,0)\wavelin(0.5,-1)}
\mov(1.65,0.1){\lin(0.5,0)}
\mov(2.05,1.1){\lin(0.5,0)\wavelin(-0.5,-1)}
\mov(1.9,0.1){\lin(0.5,0)}
\mov(0.7,0.6){$g_{aa}$}
\mov(1.65,0.6){$g_{aa}$}
\mov(-0.4,0.9){$g_{ab}$}
\mov(-1.3,0){$b$}
\mov(-1.4,1){$b$}
\mov(1.7,0){$a$}
\mov(1.6,1){$a$}
}
\GRAPH(hsize=1) 
{}

\GRAPH(hsize=3) 
{
\Nhalfperiods=10
\mov(-0.4,0.5){\large$16$}
\mov(0.7,0.6){\wavelin(0.5,0)\lin(-0.5,0.5)\lin(-0.5,-0.5)}
\mov(1.05,0.6){\arcto(0.9,0.5)[+0.5]\arcto(0.9,-0.5)[-0.5]}
\mov(1.8,1.1){\wavelin(0.5,0)\lin(0.25,-0.5)}
\mov(1.65,0.1){\lin(0.5,0)}
\mov(2.05,1.1){\lin(0.5,0)\lin(-0.25,-0.5)}\mov(1.77,0.6){\wavelin(0,-0.5)}
\mov(1.89,0.1){\lin(0.5,0)}
\mov(0.7,0.6){$g_{aa}$}
\mov(1.65,0.6){$g_{aa}$}
\mov(-0.4,0.9){$g_{ab}$}
\mov(-1.3,0){$b$}
\mov(-1.4,1){$b$}
\mov(1.7,0){$a$}
\mov(1.6,1){$a$}
}
} 
\vspace{0.3cm}

\hbox{          

\GRAPH(hsize=1){}


\GRAPH(hsize=1) 
{}

\GRAPH(hsize=3) 
{
\Nhalfperiods=10
\mov(-0.5,0.5){\large$8p$}
\mov(0.7,0.6){\wavelin(0.5,0)\lin(-0.5,0.5)\lin(-0.5,-0.5)}
\mov(1.05,0.6){\lin(0.9,0.5)\lin(0.9,-0.5)}
\mov(1.8,1.1){\wavelin(0.5,0)\lin(0,-1)}
\mov(1.65,0.1){\wavelin(0.5,0)}
\mov(2.05,1.1){\lin(0.5,0)\lin(0,-1)}
\mov(1.9,0.1){\lin(0.5,0)}
\mov(1.3,-0.2){$g_{cb}$}
\mov(1.2,1.3){$g_{cb}$}
\mov(-0.4,0.9){$g_{ac}$}
\mov(-1.3,0){$a$}
\mov(-1.4,1){$a$}
\mov(1.7,0){$b$}
\mov(1.6,1){$b$}
}
\GRAPH(hsize=1) 
{}

\GRAPH(hsize=3) 
{
\Nhalfperiods=10
\mov(-0.1,0.5){\large$8p$}
\mov(0.8,1.1){\lin(2.0,0)}
\mov(1.2,1.1){\wavelin(0,-1)}
\mov(2.0,1.1){\wavelin(0,-0.3)}
\mov(1.85,0.6){\Circle(0.4)}
\mov(1.7,0.4){\wavelin(0,-0.3)}
\mov(0.1,0.1){\lin(2.0,0)}
\mov(-0.1,0.6){$g_{ab}$}
\mov(1.55,0.85){$g_{ac}$}
\mov(1.4,0.3){$g_{ac}$}
\mov(-0.4,-0.35){$b$}
\mov(-0.5,1.3){$a$}
\mov(1.15,-0.35){$b$}
\mov(0.95,1.3){$a$}
}
\GRAPH(hsize=1) 
{}

\GRAPH(hsize=3) 
{
\Nhalfperiods=10
\mov(-0.3,0.5){\large$4p$}
\Nhalfperiods=10
\mov(1.1,0.6){\wavelin(0.5,0)\lin(-0.5,0.5)\lin(-0.5,-0.5)}
\mov(1.45,0.6){\arcto(0.9,0.5)[+0.5]\arcto(0.9,-0.5)[-0.5]}
\mov(2.2,1.1){\lin(0.5,0)}
\mov(2.0,1.1){\wavelin(0,-0.3)}
\mov(1.85,0.6){\Circle(0.4)}
\mov(1.7,0.4){\wavelin(0,-0.3)}
\mov(1.65,0.1){\lin(0.5,0)}
\mov(1.65,0.85){$g_{ac}$}
\mov(1.5,0.3){$g_{ac}$}
\mov(-0.9,0){$b$}
\mov(-1.0,1){$b$}
\mov(1.55,0){$a$}
\mov(1.45,1){$a$}
\mov(-0.65,0.8){$g_{ab}$}
}
} 
\vspace{0.3cm}

\hbox{          

\GRAPH(hsize=1){}


\GRAPH(hsize=1) 
{}

\GRAPH(hsize=3) 
{
\Nhalfperiods=10
\mov(-0.6,0.5){\large$2p$}
\mov(0.4,0.6){\wavelin(0.5,0)\lin(-0.5,0.5)\lin(-0.5,-0.5)}
\mov(1.3,0.6){\Circle(1)\wavelin(0,0.5)\wavelin(0,-0.5)}
\mov(2.15,0.6){\wavelin(-0.5,0)\lin(0.5,0.5)\lin(0.5,-0.5)}
\mov(0,1){$g_{ac}$}
\mov(1.5,1){$g_{cb}$}
\mov(-1.1,0){$a$}
\mov(-1.2,1){$a$}
\mov(2.05,0){$b$}
\mov(1.95,1){$b$}
\mov(0,-0.2){$g_{cc}$}
}
\GRAPH(hsize=1) 
{}

\GRAPH(hsize=3) 
{
\Nhalfperiods=10
\mov(-0.4,0.5){\large$4p$}
\mov(0.6,0.6){\wavelin(0.5,0)\lin(-0.5,0.5)\lin(-0.5,-0.5)}
\mov(1.3,0.6){\Circle(0.6)}
\mov(1.95,0.6){\wavelin(-0.5,0)\arcto(0.8,0.5)[+0.5]\arcto(0.8,-0.5)[-0.5]}
\mov(2.3,1){\wavelin(0,-0.85)}
\mov(0.1,0.85){$g_{bc}$}
\mov(1.1,0.85){$g_{ac}$}
\mov(2.0,0.6){$g_{aa}$}
\mov(-1.15,0){$b$}
\mov(-1.25,1){$b$}
\mov(1.85,0){$a$}
\mov(1.75,1){$a$}
}
\GRAPH(hsize=1) 
{}

\GRAPH(hsize=3) 
{
\Nhalfperiods=10
\mov(-0.4,0.5){\large$p^2$}
\mov(0.6,0.6){\wavelin(0.5,0)\lin(-0.5,0.5)\lin(-0.5,-0.5)}
\mov(1.25,0.6){\Circle(0.6)}
\mov(1.4,0.6){\wavelin(0.35,0)}
\mov(1.9,0.6){\Circle(0.6)}
\mov(2.55,0.6){\wavelin(-0.5,0)\lin(0.5,0.5)\lin(0.5,-0.5)}
\mov(-0.1,0.85){$g_{ac}$}
\mov(1.85,0.85){$g_{db}$}
\mov(0.65,0.15){$g_{cd}$}
\mov(-1.25,0){$a$}
\mov(-1.38,1){$a$}
\mov(2.27,0){$b$}
\mov(2.13,1){$b$}
}
} 
\vspace{0.7cm}
}
\Lengthunit=1cm
\Nhalfperiods=7
\vbox{

\hbox{          
\GRAPH(hsize=1) 
{}

{
}
\GRAPH(hsize=1) 
{}

\GRAPH(hsize=3) 
{
\mov(1.2,0.6){\LARGE c}
}
\GRAPH(hsize=1) 
{}

\GRAPH(hsize=3) 
{
\Nhalfperiods=10
\mov(-0.4,0.5){\large$-p$}
\Linewidth{4pt}
\mov(0.17,0.6){\lin(0.5,0)}
\Linewidth{1.2pt}
\mov(0.9,0.6){\Circle(1)}
\mov(1.75,0.6){\wavelin(-0.5,0)\lin(0.5,0.5)\lin(0.5,-0.5)}
\mov(1.2,0.9){$g_{ca}$}
\Nhalfperiods=7
}
\GRAPH(hsize=1) 
{}

\GRAPH(hsize=3) 
{
\mov(-0.2,0.5){\large$-2$}
\Nhalfperiods=10
\Linewidth{4pt}
\mov(0.47,0.6){\lin(0.5,0)}
\Linewidth{1.2pt}
\mov(0.65,0.6){\arcto(0.9,0.5)[+0.5]\arcto(0.9,-0.5)[-0.5]}
\mov(1.4,1.1){\wavelin(0,-1)\lin(0.5,0)}
\mov(1.25,0.1){\lin(0.5,0)}
\mov(1.3,0.6){$g_{aa}$}
\Nhalfperiods=7
}
} 
\vspace{0.3cm}

\hbox{          

\GRAPH(hsize=1){}


\GRAPH(hsize=1) 
{}

\GRAPH(hsize=3) 
{
\Nhalfperiods=10
\mov(-0.7,0.5){\large$2p$}
\Nhalfperiods=10
\Linewidth{4pt}
\mov(-0.2,0.6){\lin(0.5,0)}
\Linewidth{1.2pt}
\mov(0.5,0.6){\Circle(1.0)}
\mov(0.85,0.6){\wavelin(0.4,0)}
\mov(1.1,0.6){\arcto(0.9,0.5)[+0.5]\arcto(0.9,-0.5)[-0.5]}
\mov(1.85,1.1){\wavelin(0,-1)\lin(0.5,0)}
\mov(1.7,0.1){\lin(0.5,0)}
\mov(1.75,0.6){$g_{aa}$}
\mov(0.2,0.95){$g_{ca}$}
\Nhalfperiods=7
}
\GRAPH(hsize=1) 
{}

\GRAPH(hsize=3) 
{
\Nhalfperiods=10
\mov(0.0,0.5){\large$4$}
\Linewidth{4pt}
\mov(0.3,0.6){\lin(0.5,0)}\Linewidth{1.2pt}
\mov(0.65,0.6){\arcto(0.9,0.5)[+0.5]\arcto(0.9,-0.5)[-0.5]}
\mov(1.4,1.1){\lin(0.5,0)\wavelin(0,-1)}
\mov(1.25,0.1){\lin(0.5,0)}
\mov(1.63,1.1){\lin(0.5,0)\wavelin(0,-1)}
\mov(1.48,0.1){\lin(0.5,0)}
\mov(0.2,0.6){$g_{aa}$}
\mov(1.35,0.6){$g_{aa}$}
}
\GRAPH(hsize=1) 
{}

\GRAPH(hsize=3) 
{
\Nhalfperiods=10
\mov(0.0,0.5){\large$4$}
\Linewidth{4pt}
\mov(0.3,0.6){\lin(0.5,0)}\Linewidth{1.2pt}
\mov(0.65,0.6){\arcto(0.9,0.5)[+0.5]\arcto(0.9,-0.5)[-0.5]}
\mov(1.4,1.1){\lin(0.5,0)\wavelin(0.5,-1)}
\mov(1.25,0.1){\lin(0.5,0)}
\mov(1.63,1.1){\lin(0.5,0)\wavelin(-0.5,-1)}
\mov(1.47,0.1){\lin(0.5,0)}
\mov(0.3,0.6){$g_{aa}$}
\mov(1.25,0.6){$g_{aa}$}
}
} 
\vspace{0.3cm}

\hbox{          

\GRAPH(hsize=1){}


\GRAPH(hsize=1) 
{}

\GRAPH(hsize=3) 
{
\Nhalfperiods=10
\mov(-0.2,0.5){\large$8$}
\Linewidth{4pt}\mov(0.5,0.6){\lin(0.5,0)}\Linewidth{1.2pt}
\mov(0.85,0.6){\arcto(0.9,0.5)[+0.5]\arcto(0.9,-0.5)[-0.5]}
\mov(1.6,1.1){\wavelin(0.5,0)\lin(0.25,-0.5)}
\mov(1.45,0.1){\lin(0.5,0)}
\mov(1.85,1.1){\lin(0.5,0)\lin(-0.25,-0.5)}\mov(1.57,0.6){\wavelin(0,-0.5)}
\mov(1.69,0.1){\lin(0.5,0)}
\mov(0.5,0.6){$g_{aa}$}
\mov(1.45,0.6){$g_{aa}$}
}
\GRAPH(hsize=1) 
{}

\GRAPH(hsize=3) 
{
\Nhalfperiods=10
\mov(-0.5,0.5){\large$4p$}
\Linewidth{4pt}\mov(0.5,0.6){\lin(0.5,0)}\Linewidth{1.2pt}
\mov(0.85,0.6){\arcto(0.9,0.5)[+0.5]\arcto(0.9,-0.5)[-0.5]}
\mov(1.6,1.1){\wavelin(0.5,0)\lin(0,-1)}
\mov(1.45,0.1){\wavelin(0.5,0)}
\mov(1.85,1.1){\lin(0.5,0)\lin(0,-1)}
\mov(1.7,0.1){\lin(0.5,0)}
\mov(1.1,-0.2){$g_{ca}$}
\mov(1.0,1.3){$g_{ca}$}
}
\GRAPH(hsize=1) 
{}

\GRAPH(hsize=3) 
{
\Nhalfperiods=10
\mov(-0.3,0.5){\large$4p$}
\Nhalfperiods=10
\Linewidth{4pt}\mov(0.5,0.6){\lin(0.5,0)}\Linewidth{1.2pt}
\mov(0.85,0.6){\arcto(1.1,0.7)[+0.5]\arcto(1.1,-0.7)[-0.5]}
\mov(1.8,1.3){\lin(0.7,0)}
\mov(1.65,1.27){\wavelin(0,-0.4)}
\mov(1.5,0.6){\Circle(0.5)}
\mov(1.35,0.35){\wavelin(0,-0.4)}
\mov(1.25,-0.1){\lin(0.7,0)}
\mov(1.25,1.0){$g_{ac}$}
\mov(1.1,0.15){$g_{ac}$}
}
} 
\vspace{0.3cm}

\hbox{          

\GRAPH(hsize=1){}


\GRAPH(hsize=1) 
{}

\GRAPH(hsize=3) 
{
\Nhalfperiods=10
\mov(-0.3,0.5){\large$p$}
\Linewidth{4pt}\mov(0.4,0.6){\lin(0.5,0)}\Linewidth{1.2pt}
\mov(1.28,0.6){\Circle(1)\wavelin(0,0.5)\wavelin(0,-0.5)}
\mov(2.15,0.6){\wavelin(-0.5,0)\lin(0.5,0.5)\lin(0.5,-0.5)}
\mov(1.6,0.8){$g_{ca}$}
\mov(0.6,-0.2){$g_{cc}$}
}
\GRAPH(hsize=1) 
{}

\GRAPH(hsize=3) 
{
\Nhalfperiods=10
\mov(-0.4,0.5){\large$p^2$}
\Linewidth{4pt}\mov(0.37,0.6){\lin(0.5,0)}\Linewidth{1.2pt}
\mov(1.25,0.6){\Circle(1.0)}
\mov(1.6,0.6){\wavelin(0.55,0)}
\mov(2.5,0.6){\Circle(1.0)}
\mov(3.35,0.6){\wavelin(-0.5,0)\lin(0.5,0.5)\lin(0.5,-0.5)}
\mov(2.8,0.85){$g_{ca}$}
\mov(1.1,0.2){$g_{dc}$}
}
} 
}
}
\label{RSBfig}
\caption{
The diagram representation of contributions to the two-point $\Gamma^{(2)}$ (a),
four-point $\Gamma^{(2,1)}_{aa}$ (b), and with the inclusion $({\phi}_{i}^{a})^{2}$
 (c) vertex functions in the one- and two-loop approximations with the corresponding weighting coefficients.
}
\end{figure}

The renormalization group description of the model
specified by the replica Hamiltonian (\ref{Hrepl}) was carried out
in the framework of the field-theoretic approach in the
two-loop approximation for systems of an arbitrary
dimension in the range from three to four. Possible
types of critical behavior and their stability in the fluctuation
domain are determined by the renormalization
group equation for the coefficients of the matrix $g_{ab}$.
They were determined by the conventional method
based on the Feynman diagram technique for the vertex
parts of the irreducible Grin functions and the renormalization
procedure. For example, in the two-loop
approximation, the two-point vertex function $\Gamma^{(2)}$, the
four-point vertex functions $\Gamma^{(4)}_{ab}$, and the two-point
function $\Gamma^{(2,1)}_{aa}$ with the insertion $({\phi}_{i}^{a})^{2}$ have the form
\begin{eqnarray}
\left.\frac{\partial\Gamma^{(2)}}{\partial k^2}\right|_{k^2=0}&=&
1+4fg_{aa}^2+2pf\sum\limits_{c=1}^{n}g_{ac}g_{ca},   \\
\left.\Gamma^{(4)}_{ab}\right|_{k_i=0}&=&g_{ab}-p\sum\limits_{c=1}^{n}g_{ac}g_{cb}
-4g_{aa}g_{ab}-4g_{ab}^2+(8+16h)g_{ab}^3+(24+8h)g_{aa}^2g_{ab}+ \nonumber\\
&+&48hg_{aa}g_{ab}^2+4g_{aa}g_{bb}g_{ab}
+8ph\sum\limits_{c=1}^{n}g_{ac}g_{cb}^2
+8phg_{ab}\sum\limits_{c=1}^{n}g_{ac}g_{cb}
+4phg_{ab}\sum\limits_{c=1}^{n}g_{ac}^2+  \nonumber \\
&+&2p\sum\limits_{c=1}^{n}g_{ac}g_{cc}g_{cb}
+4pg_{aa}\sum\limits_{c=1}^{n}g_{ac}g_{cb}
+p^2\sum\limits_{c,d=1}^{n}g_{ac}g_{cd}g_{db}, \\  \nonumber
\left.\Gamma^{(2,1)}_{aa}\right|_{k_i=0}&=&1-p\sum\limits_{c=1}^{n}g_{ca}
-2g_{aa}+2pg_{aa}\sum\limits_{c=1}^{n}g_{ca}+(4+12h)g_{aa}^2+ \\
&+&6ph\sum\limits_{c=1}^{n}g_{ca}^2+p\sum\limits_{c=1}^{n}g_{cc}g_{ca}
+p^2\sum\limits_{c,d=1}^{n}g_{dc}g_{ca}, \\  \nonumber
\end{eqnarray}
where the notation
\begin{eqnarray}
&&\left. f(d)=-\frac{1}{J^2}\frac{\partial}{\partial k^2}\int
\frac{d^{d}k_1d^{d}k_2}{(k_1^2+1)(k_2^2+1)((k_1+k_2+k)^2+1)}\right|_{k^2=0},  \\ \nonumber
&&h(d)=\frac{1}{J^2}\int
\frac{d^{d}k_1d^{d}k_2}{(k_1^2+1)^2(k_2^2+1)((k_1+k_2)^2+1)}, \  \
J = \int d^{d}k/(k^2+1)^2, \\  \nonumber
\end{eqnarray}
is used, and the redefinition $g_{ab}\rightarrow g_{ab}/J$ is carried out.
The diagram representation of the corresponding contributions
to $\Gamma^{(2)}$, $\Gamma^{(4)}_{ab}$ and
$\Gamma^{(2,1)}_{aa}$ is shown in \ref{RSBfig}.

However, the subsequent renormalization procedure
for the vertex functions and the calculation of the $\beta$ and $\gamma$
 functions, which determine the renormalization group
transformations for the interaction constants, are difficult
due to the complicated structure of relations (\ref{mrule}) and
(\ref{mrule2}) defining operations with matrices $g_{ab}$. The step-like
structure of the function $g(x)$ established in \cite{4,5,6}
makes it possible to implement the renormalization
procedure. In this paper, we restrict ourselves to the
consideration of the one-step function $g(x)$:
\begin{equation}
 g(x)= \left\{ \begin{array}{c}
  g_0,  0 \leq x < x_0, \\ \displaystyle
  g_1,  x_0 < x \leq 1,
 \end{array} \right.
\end{equation}
where the coordinate of the step $0\leq x_0 \leq 1$ is an arbitrary
parameter that does not evolve under scale transformations
and remains the same as in the seed function
$g_0(x)$. As a result, the renormalization group transformations
of the replica Hamiltonian with RSB are determined
by the three parameters $\tilde{g}, g_0, g_1$.

The critical properties of the model can be revealed by analyzing the coefficients
$\beta_i(\tilde{g}, g_0, g_1)$ $(i=1,2,3)$, $\gamma_{\phi}(\tilde{g}, g_0, g_1)$, and
$\gamma_{\phi^2}(\tilde{g}, g_0, g_1)$ of the renormalization group Callan-Symanzik
equation \cite{14}. We obtained the following expressions for the $\beta$- and $\gamma$
functions in the two-loop approximation in the form of series in the renormalized
parameters $\tilde{g}$,$g_0$ and $g_1$:
\begin{eqnarray} \label{beta}
\displaystyle \beta_1&=&-\tilde{g}+\left (8+p\right ){\tilde{g}}^{2}
-p{\it x_0}\,{g_0}^{2}-p\left (1-{\it x_0}\right ){g_1}^{2}+[(8\,f
-40\,h+20)p \nonumber \\
\displaystyle &+&16\,f-176\,h+88]{\tilde{g}}^{3}+\left (24\,h-8\,f-12
\right ){\it x_0}\,p\tilde{g}{g_0}^{2}+(24\,h-8\,f \nonumber \\
\displaystyle &-&12)\left (1-{\it x_0}\right )p\tilde{g}{g_1}^{2}-\left (16\,h-8\right )
{\it x_0}\,p{g_0}^{3}-\left (16\,h-8\right )\left (1-{\it x_0}
\right )p{g_1}^{3}, \nonumber \\
 \displaystyle\beta_2&=&-g_0+\left (4+2\,p\right )\tilde{g}g_0+
\left (2\,p{\it x_0}-4\right ){g_0}^{2}+2\,\left (1-{\it x_0}\right )pg_0g_1 \nonumber \\
\displaystyle &+&[(8\,f-48\,h+28)p+16\,f-48\,h+24]{\tilde{g}}^{2}g_0
-[((32\,h-16){\it x_0} \\
\displaystyle &+&8-32\,h )p+48-96\,h ]\tilde{g}{g_0}^{2}
-\left (32\,h-16\right)\left (1-{\it x_0}\right )p\tilde{g}g_0g_1 \nonumber \\
\displaystyle &+&[\left (48\,h-8\,f-20\right ){\it x_0}\,p
-32\,h+16]{g_0}^{3}+\left (32\,h-8\right )\left (1-{\it x_0}\right )p{g_0}^{2}g_1  \nonumber \\
\displaystyle &+&\left (16\,h-12-8\,f\right )\left (1-{\it x_0}\right )pg_0{g_1}^{2},  \nonumber \\
 \displaystyle\beta_3&=&-g_1+p{\it x_0}\,{g_0}^{2}-\left [p\left
({\it x_0}-2\right )+4\right ]{g_1}^{2}+\left (4+2\,p\right )\tilde{g}g_1
+[(8\,f-48\,h \nonumber \\
\displaystyle &+&28 )p+16\,f-48\,h+24]g_1{\tilde{g}}^{2}-\left (16\,h-8\right )
{\it x_0}\,p\tilde{g}{g_0}^{2}-[ (\left (8-16\,h\right ) {\it x_0} \nonumber \\
\displaystyle &-&8 )p+48-96\,h]u_{{0
}}{g_1}^{2}+\left (16\,h-8\right ){\it x_0}\,p{g_0}^{3}+\left
(8\,h-8\,f-4\right ){\it x_0}\,pg_1{g_0}^{2}\nonumber \\
\displaystyle &+&\left [\left (8\,f-24\,h+12\right ){\it x_0}\,p+\left (48\,h-8\,f
-20\right )p+16-32\,h\right ]{g_1}^{3}, \nonumber \\
\displaystyle\gamma_{\phi}&=&4(4-d)\,f(d)\left [ (p+2){\tilde{g}}^{2}
-p{\it x_0}{g_0}^{2}-p(1-{\it x_0}){g_1}^{2}\right ], \nonumber \\
\displaystyle\gamma_{{\phi}^{2}}&=&-(4-d)\left [ (p+2)\tilde{g}+p{\it x_0}{g_0}
+p(1-{\it x_0})g_1-2(6\,h-2\,f-3\right )\left ((p+2\right ){\tilde{g}}^{2}\nonumber \\
\displaystyle&-&p{\it x_0}{g_0}^{2}-p\left (1-{\it x_0}){g_1}^{2}) \right ]. \nonumber
\end{eqnarray}
In order to compare the results of this paper with those
obtained in \cite{4,5,6}, we reversed, by analogy with \cite{4,5,6},
the signs of the off-diagonal elements in the matrix:
$g_{a\neq b}\rightarrow - g_{a\neq b}$. As a result, $g_0$ and $g_1$ became positive
definite. The integrals $f(d)$ and $h(d)$ were calculated
numerically for $3\leq d <4 $.

It is well known that the series used in perturbation theory are asymptotic, and vertices
of the interaction of the order parameter fluctuations in the fluctuation domain $\tau
\rightarrow 0$ are sufficiently large to directly apply expressions (\ref{beta}). For
this reason, in order to extract the physical information from those expressions, we use
the generalized (for the three-parameter case) Pade-Borel method for summing asymptotic
series. The direct and inverse Borel transformations have the form
\begin{equation}
\begin{array}{rl}
  & f(\tilde{g},g_0,g_1)=\sum\limits_{i,j,k}c_{ijk}\tilde{g}^i g_0^j g_1^k=\int\limits_{0}^{\infty}e^{-t}F(\tilde{g}t,g_0t,g_1t)dt,  \\
  & F(\tilde{g},g_0,g_1)=\sum\limits_{i,j,k}\frac{c_{ijk}}{(i+j+k)!}\tilde{g}^i g_0^j g_1^k.
\end{array}
\end{equation}
In order to find the analytic continuation of the Borel
image of a function, we use the following series in the
auxiliary variable $\theta$
\begin{equation}
   {\tilde{F}}(\tilde{g},g_0,g_1,\theta)=\sum\limits_{k=0}^{\infty}\theta^k\sum\limits_{i=0}^{k}\sum\limits_{j=0}^{k-i}\frac{c_{i,j,k-i-j}}{k!}\tilde{g}^i g_0^j g_1^{k-i-j},
\end{equation}
The Pade approximation [L/M] is applied to this
series at the point $\theta=1$. This technique was
successfully used in \cite{8} to describe the critical behavior of certain
systems characterized by several interaction vertices
of the order parameter fluctuations. The symmetry
conservation property of a system when applying a
Pade approximant in the variable . becomes essential
in the description of multivertex models. In this paper,
we used the [2/1] approximant for the calculation of
$\beta$-functions in the two-loop approximation.

\section{Calculation results}

It is well known that the nature of the critical behavior
is determined by the existence of a stable fixed point
satisfying the system of equations
\begin{equation}
\label{beta_sys}
\beta_{i}(\tilde{g}^*,g_0^*,g_1^*)=0\ \ \ \ \ \ \ \ \ \ \ \ \ \   (i=1,2,3).
\end{equation}

By numerically solving system (\ref{beta_sys}) for functions
found by the Pade Borel summation technique (for
$p=1,2,3$ ), three types of nontrivial fixed points
were found in the physically interesting domain of
parameters $\tilde{g}^*,g_0^*,g_1^* \geq 0$ (see Tables 1-3). The firsttype
fixed point with $\tilde{g}^*\neq 0, g_0^* = g_1^* = 0$ corresponds
to the critical behavior of the homogeneous system;
the second-type fixed point with $\tilde{g}^*\neq 0$ and $g_0^* = g_1^* \neq 0$
corresponds to the critical behavior of the disordered
system with replica symmetry; and the thirdtype
fixed point with $\tilde{g}^*\neq 0, g_0^* = 0$  and $g_1^* \neq 0$ corresponds
to the critical behavior of the disordered system
with RSB. The values of and at the fixed point
with RSB depend on the coordinate of the step $x_0$.
Tables 1-3 present the values of and for $0\leq x_0 \leq 1$
with the step $\Delta x_0 =0,1$
The possibility to realize one or another type of the
critical behavior for each $p$ depends on the stability of
the corresponding fixed point. The stability criterion of
a fixed point reduces to a condition that the eigenvalues
$\lambda_i$ of the matrix
\begin{equation}
B_{i,j}=\frac{\partial\beta_i(\tilde{g}^*,g_0^*,g_1^*)}{\partial{g_j}}
\end{equation}
belong to the complex right half-plane.

\begin{table}[p]
\hspace{127mm} {\bf Table 1}
\vspace{-6mm}
\begin{center}
{Coordinates of the FPs and eigenvalues of the stability matrix for $p=1$: \\
 \vspace{-2mm} a) dimension $d=3.0$ \vspace{2mm}}
\begin{tabular}{|c|c|c|c|c|c|c|} \hline
  Type&$x_0$& $\tilde{g}^*$ & $g_{0}^*$ & $g_{1}^*$   & $\lambda_1$\hspace{1cm} $\lambda_2$ & $\lambda_3$ \\  \hline
   1  &     &  0.1774       &      0    &      0      &  0.6536 \hspace{5mm}    $-0.1692$   &  $-0.1692$ \\  \hline
   2  &     &  0.1844       & $0.0812$  &  $0.0812$   & $0.5253\pm0.0893i$                  &  0.2112      \\  \hline
   3  & 0.0 &  0.1844       &      0    &  $0.0812$   & $0.5253\pm0.0893i$              & $-0.0392$  \\
      & 0.1 &  0.1840       &      0    &  $0.0829$   & $0.5352\pm0.0983i$              & $-0.0492$  \\
      & 0.2 &  0.1835       &      0    &  $0.0846$   & $0.5471\pm0.1067i$              & $-0.0599$  \\
      & 0.3 &  0.1830       &      0    &  $0.0863$   & $0.5607\pm0.1133i$              & $-0.0712$  \\
      & 0.4 &  0.1824       &      0    &  $0.0880$   & $0.5765\pm0.1180i$              & $-0.0832$  \\
      & 0.5 &  0.1817       &      0    &  $0.0895$   & $0.5951\pm0.1203i$              & $-0.0959$  \\
      & 0.6 &  0.1810       &      0    &  $0.0910$   & $0.6172\pm0.1189i$              & $-0.1093$  \\
      & 0.7 &  0.1802       &      0    &  $0.0924$   & $0.6439\pm0.1114i$              & $-0.1234$  \\
      & 0.8 &  0.1793       &      0    &  $0.0936$   & $0.6760\pm0.0921i$              & $-0.1381$  \\
      & 0.9 &  0.1784       &      0    &  $0.0947$   & $0.7135\pm0.0353i$              & $-0.1534$  \\
      & 1.0 &  0.1774       &      0    &  $0.0957$   & 0.8573 \hspace{5mm} 0.6536      & $-0.1692$  \\  \hline
\end{tabular} \end{center}
\begin{center}
{\vspace{-2mm} b) dimension $d=3.985$ \\ \vspace{2mm}}
\begin{tabular}{|c|c|c|c|c|c|c|} \hline
  Type&$x_0$& $\tilde{g}^*$ & $g_{0}^*$ & $g_{1}^*$ & $\lambda_1$\hspace{1cm}$\lambda_2$ & $\lambda_3$ \\  \hline
   1  &     & $0.0917    $ &      0    &      0     & $0.6315$\hspace{5mm}   $-0.4163$   & $-0.4163$  \\  \hline
   2  &     & $0.1231    $ &  $0.1090$ &   $0.1090$ & $0.6986\pm0.1311i$                  & $ 0.0022$  \\  \hline
   3  & 0.0 & $0.1231    $ &      0    &   $0.1090$ & $0.7047\pm0.1069i$                  & $-0.0363$  \\  \hline
\end{tabular} \end{center}
\begin{center}
{\vspace{-2mm} c) dimension $d=3.986$ \\ \vspace{2mm}}
\begin{tabular}{|c|c|c|c|c|c|c|} \hline
  Type&$x_0$& $\tilde{g}^*$ & $g_{0}^*$ & $g_{1}^*$  & $\lambda_1$\hspace{1cm} $\lambda_2$ & $\lambda_3$ \\  \hline
   1  &     & $0.0916    $ &      0    &      0      & $0.6318$\hspace{5mm}   $-0.4165$   & $-0.4165$  \\  \hline
   2  &     & $0.1230    $ &  $0.1092$ &   $0.1092$  & $0.6895\pm0.1453i$                  & $-0.0076$  \\  \hline
   3  & 0.0 & $0.1230    $ &      0    &   $0.1092$  & $0.7018\pm0.0935i$                  & $-0.0359$  \\  \hline
\end{tabular} \end{center}
\end{table}
\begin{table}[p]
\hspace{127mm} {\bf Table 2}
\vspace{-6mm}
\begin{center}
{ Coordinates of the FPs and eigenvalues of the stability matrix for $p=2$:\\
  \vspace{-2mm} a) dimension $d=3.0$ \vspace{2mm}}
\begin{tabular}{|c|c|c|c|c|c|c|c|} \hline
  Type&$x_0$& $\tilde{g}^*$ & $g_{0}^*$ & $g_{1}^*$    &$\lambda_1$     &    $\lambda_2$      &   $\lambda_3$  \\  \hline
   1  &     & 0.155830     &    0       &     0        & 0.667315       &  $-0.001672$        &  $-0.001672$   \\  \hline
   2  &     & 0.155831     & $0.000584$ &  $0.000584$  & 0.667312       &  0.001682           &  $0.000004$    \\  \hline
   3  & 0.0 & 0.155831     &    0       &  $0.000584$  & 0.667313       &  0.001683           &  $-0.000001$   \\
      & 0.1 & 0.155831     &    0       &  $0.000614$  & 0.667313       &  0.001684           &  $-0.000088$   \\
      & 0.2 & 0.155831     &    0       &  $0.000648$  & 0.667313       &  0.001685           &  $-0.000186$   \\
      & 0.3 & 0.155831     &    0       &  $0.000686$  & 0.667313       &  0.001686           &  $-0.000296$   \\
      & 0.4 & 0.155831     &    0       &  $0.000729$  & 0.667313       &  0.001687           &  $-0.000419$   \\
      & 0.5 & 0.155831     &    0       &  $0.000778$  & 0.667313       &  0.001687           &  $-0.000559$   \\
      & 0.6 & 0.155831     &    0       &  $0.000833$  & 0.667313       &  0.001688           &  $-0.000717$   \\
      & 0.7 & 0.155831     &    0       &  $0.000896$  & 0.667314       &  0.001690           &  $-0.000901$   \\
      & 0.8 & 0.155831     &    0       &  $0.000971$  & 0.667314       &  0.001692           &  $-0.001116$   \\
      & 0.9 & 0.155831     &    0       &  $0.001058$  & 0.667315       &  0.001694           &  $-0.001369$   \\
      & 1.0 & 0.155830     &    0       &  $0.001163$  & 0.667316       &  0.001696           &  $-0.001672$   \\   \hline
\end{tabular} \end{center}
\begin{center}
{ \vspace{-2mm} b) dimension $d=3.10$ \\ \vspace{2mm}}
\begin{tabular}{|c|c|c|c|c|c|c|c|} \hline
  Type&$x_0$&$\tilde{g}^*$& $g_{0}^*$ & $g_{1}^*$ & $\lambda_1$  &  $\lambda_2$  & $\lambda_3$ \\  \hline
   1  &     & $0.1499955$  &      0    &      0   & $ 0.689608$  &  $-0.009539$  & $-0.009539$  \\  \hline
   2  &     & $0.1500170$  & $0.00325$& $0.00325$ & $ 0.689535$  &  $ 0.009887$  & $-0.000003$  \\  \hline
   3  & 0.0 & $0.1500170$  &      0   & $0.00325$ & $ 0.689535$  &  $ 0.009887$  & $ 0.000109$  \\
      & 0.1 & $0.1500169$  &      0   & $0.00341$ & $ 0.689535$  &  $ 0.009899$  & $-0.000401$  \\
      & 0.2 & $0.1500167$  &      0   & $0.00360$ & $ 0.689536$  &  $ 0.009926$  & $-0.000961$  \\ \hline
\end{tabular}\end{center}
\begin{center}
{ \vspace{-2mm} c) dimension $d=3.999$ \\ \vspace{2mm}}
\begin{tabular}{|c|c|c|c|c|c|c|c|} \hline
  Type&$x_0$&$\tilde{g}^*$& $g_{0}^*$ & $g_{1}^*$  & $\lambda_1$  &  $\lambda_2$  & $\lambda_3$ \\  \hline
   1  &     & $0.089762$  &      0    &      0     & $1.119442$  &  $-0.133591$ & $-0.133591$  \\  \hline
   2  &     & $0.092307$  & $0.036991$& $0.036991$ & $1.103421$  &  $ 0.227335$ & $-0.025378$  \\  \hline
   3  & 0.0 & $0.092307$  &      0    & $0.036991$ & $1.103421$  &  $0.227335$  & $0.030783 $  \\
      & 0.1 & $0.092270$  &      0    & $0.038723$ & $1.102142$  &  $0.235506$  & $0.021563 $  \\
      & 0.2 & $0.092205$  &      0    & $0.040559$ & $1.100913$  &  $0.244667$  & $0.011135 $  \\
      & 0.3 & $0.092108$  &      0    & $0.042500$ & $1.099845$  &  $0.254810$  & $-0.000648$  \\
      & 0.4 & $0.091970$  &      0    & $0.044547$ & $1.099106$  &  $0.265820$  & $-0.013939$  \\ \hline
\end{tabular} \end{center} \end{table}
\begin{table}[p]
\hspace{127mm} {\bf Table 3}
\vspace{-6mm}
\begin{center}
{Coordinates of the FPs and eigenvalues of the stability matrix for $p=3$:\\
 \vspace{-2mm} a) dimension $d=3.0$ \vspace{2mm}}
\begin{tabular}{|c|c|c|c|c|c|c|c|} \hline
  Type &$x_0$& $\tilde{g}^*$ & $g_{0}^*$ & $g_{1}^*$   & $\lambda_1$  &  $\lambda_2$     & $\lambda_3$    \\  \hline
   1  &     & 0.1383     &    0         &     0       & 0.6814       &  0.1315          & 0.1315       \\  \hline
   2  &     & 0.1419     & $-0.0359$    & $-0.0359$   & 0.6727       & $-0.0891$        & $-0.1450$    \\  \hline
   3  & 0.0 & 0.1419     &    0         & $-0.0359$   & 0.6727       & $-0.0891$        & $-0.0058$    \\
      & 0.1 & 0.1420     &    0         & $-0.0382$   & 0.6727       & $-0.0865$        & 0.0011       \\
      & 0.2 & 0.1420     &    0         & $-0.0408$   & 0.6728       & $-0.0836$        & 0.0088       \\
      & 0.3 & 0.1421     &    0         & $-0.0439$   & 0.6730       & $-0.0802$        & 0.0175       \\
      & 0.4 & 0.1420     &    0         & $-0.0474$   & 0.6734       & $-0.0764$        & 0.0273       \\
      & 0.5 & 0.1420     &    0         & $-0.0516$   & 0.6738       & $-0.0719$        & 0.0385       \\
      & 0.6 & 0.1418     &    0         & $-0.0565$   & 0.6745       & $-0.0668$        & 0.0515       \\
      & 0.7 & 0.1415     &    0         & $-0.0625$   & 0.6755       & $-0.0606$        & 0.0667       \\
      & 0.8 & 0.1409     &    0         & $-0.0699$   & 0.6768       & $-0.0533$        & 0.0845       \\
      & 0.9 & 0.1400     &    0         & $-0.0793$   & 0.6787       & $-0.0443$        & 0.1058       \\
      & 1.0 & 0.1383     &    0         & $-0.0915$   & 0.6814       & $-0.0331$        & 0.1315       \\   \hline
\end{tabular} \end{center}
\begin{center}
{\vspace{-2mm} b) dimension $d=3.999$ \\ \vspace{2mm}}
\begin{tabular}{|c|c|c|c|c|c|c|c|} \hline
  Type &$x_0$& $\tilde{g}^*$ & $g_{0}^*$ & $g_{1}^*$ & $\lambda_1$ &  $\lambda_2$     & $\lambda_3$    \\  \hline
   1  &     & $0.081989$ &    0        &    0       & $ 1.113633$ & $-0.000820$ & $-0.000820$    \\  \hline
   2  &     & $0.081989$ & $0.000171$  & $0.000171$ & $1.113633$  & $0.000822$  & $-0.000228$    \\  \hline
   3  & 0.0 & $0.081989$ &    0        & $0.000171$ & $1.113633$  & $0.000822$  & $0.000228 $   \\
      & 0.1 & $0.081989$ &    0        & $0.000183$ & $1.113633$  & $0.000822$  & $0.000188 $   \\
      & 0.2 & $0.081989$ &    0        & $0.000196$ & $1.113633$  & $0.000823$  & $0.000142 $   \\
      & 0.3 & $0.081989$ &    0        & $0.000212$ & $1.113633$  & $0.000823$  & $0.000088 $   \\
      & 0.4 & $0.081989$ &    0        & $0.000230$ & $1.113633$  & $0.000823$  & $0.000025 $   \\
      & 0.5 & $0.081989$ &    0        & $0.000251$ & $1.113633$  & $0.000823$  & $-0.000050$   \\
      & 0.6 & $0.081989$ &    0        & $0.000277$ & $1.113633$  & $0.000824$  & $-0.000140$   \\
      & 0.7 & $0.081989$ &    0        & $0.000309$ & $1.113633$  & $0.000824$  & $-0.000251$   \\
      & 0.8 & $0.081989$ &    0        & $0.000350$ & $1.113633$  & $0.000825$  & $-0.000391$   \\
      & 0.9 & $0.081989$ &    0        & $0.000402$ & $1.113633$  & $0.000826$  & $-0.000574$   \\
      & 1.0 & $0.081989$ &    0        & $0.000473$ & $1.113633$  & $0.000828$  & $-0.000820$   \\   \hline
\end{tabular} \end{center}\end{table}

An analysis of
$\lambda_i$ for every type of the fixed point (see Tables 1-3) provides
the following conclusions.

1) For the three-dimensional Ising model ($p = 1$),
the second-type fixed point is stable (Table 1(a)). The
complex values $\lambda_1$ and $\lambda_2$ for positive
$|\lambda_1|,|\lambda_2|$ and $\lambda_3$,
show that the second-type fixed point, in contrast to the
third-type one, is a stable focus in the parametric space
$(\tilde{g},g_0,g_1)$, and the renormalization group flows
approach the second-type fixed point by a spiral curve.
At the threshold dimension $d_c = 3.986$ (see Table 1(b),
(c)), the second-type fixed point looses stability ($\lambda_3$
changes sign). Since all other fixed points remain unstable
in the entire range of the dimension variation $3\leq d<4$,
 no critical behavior is realized in the system at
$3.986\leq d$ due to the replica symmetry breaking. The
analysis of the behavior of renormalization group flows
at $3.986\leq d$ provides the following results.

2) For the three-dimensional XY model ($p = 2$),
small values of $\lambda_i$ (see Table 2(a)) indicate that the second-
type replica symmetric fixed point is weakly stable.
However, already for the dimension $d_c=3.1$ (see
Table 2(b), (c)), the third-type fixed point with RSB
effects becomes stable. However, the critical behavior
determined by this point is nonuniversal and depends
on the parameter $x_0$ and, therefore, on the concentration
of impurities. A stability analysis of the third-type fixed
point reveals that it is stable only in the interval $0\leq x_0\leq x_c(d)$,
where $x_c$ is a threshold value of the parameter,
which depends on the dimension of the system. For
example, for $d=3.1$ $x_c=0.1$; and, for $d=3.999$ $x_c=0.3$
In the interval $x_c(d) < x_0 < 1$, all fixed points are
unstable.

3) For the isotropic three-dimensional Heisenberg
model ($p = 3$), the first-type fixed point becomes stable
(Table 3(a)), while at the other fixed points the constants
$g_0^*$ and $g_1^*$ take physically senseless negative
values. Only at $d_c = 3.999$, $g_0^*$ and $g_1^*$ take physically
meaningful values for the third-type point, and this
point becomes stable in the interval $0\leq x_0\leq 0.4$
(Table 3(b), (c)). In the interval $0.4< x_0 <1$, all fixed
points are unstable.

Note that although the calculations indicate the stability
of the impurity replica symmetric second-type
fixed point, there is reason to believe that, in the higher
order approximations (as is the case for disordered systems
considered without taking into account RSB
effects \cite{11}), the first-type fixed point, which corresponds
to the critical behavior of the homogeneous system,
will become stable. On the one hand, this is indicated
by the very weak stability ($\lambda_3 = 0.000004$) of the
second-type fixed point and by the fact that the threshold
value of the order parameter $p_c = 2.0114$ found in
the two-loop approximation, which separates the critical
behavior domains determined by the first- ($p > p_c$)
and second-type ($p < p_c$) fixed points, is very close to
$p = 2$. This explains a very slow variation of the eigenvalues
$\lambda_i$ of the stability matrix for the disordered XY
model with the variation of the system dimension
(Table 2). On the other hand, the negative value of the
critical heat capacity coefficient $\alpha$ of the XY model also
suggests, according to the Harris criterion, that the critical
behavior of the model is stable with respect to the
influence of the quenched disorder and, therefore, it can
be expected that $p_c < 2$ in the higher order approximations.
For example, the value $p_c = 1.912(4)$ was found
in \cite{15} on the basis of the six-loop approximation with
the use of the pseudo $\varepsilon$-expansion and the Pade Borel
Leroy summation technique with a thoroughly chosen
fitting parameter.

\begin{table}[t]
\hspace{127mm} {\bf Table 4} \vspace{-6mm}
\begin{center}
{Critical exponents of the 3D models for RS FP's}
\begin{tabular}{|lllccccc|} \hline
    Model   &  FP  &                          &  $\eta$   &   $\nu$   & $\gamma$  & $\beta$  & $\alpha$      \\ \hline
    Ising   &  RS1 &  this work               & 0.028     & 0.631     & 1.244     & 0.324    &    0.107      \\
            &      &  Ref.~\protect{\cite{16}}& 0.031(4)  & 0.630(2)  & 1.241(2)  & 0.325(2) &    0.110(5)   \\
            &  RS2 &  this work               & 0.028     & 0.672     & 1.329     & 0.345    & $-$0.015      \\
            &      &  Ref.~\protect{\cite{17}}& 0.030(3)  & 0.678(10) & 1.330(17) & 0.349(5) & $-$0.034(30)  \\ \hline
       XY   &  RS1 &  this work               & 0.029     & 0.667     & 1.318     & 0.343    & $-$0.001      \\
            &      &  Ref.~\protect{\cite{16}}& 0.034(3)  & 0.669(1)  & 1.316(1)  & 0.346(1) & $-$0.007(6)   \\ \hline
Heisenberg  &  RS1 &  this work               & 0.028     & 0.697     & 1.379     & 0.369    & $-$0.092      \\
            &      &  Ref.~\protect{\cite{16}}& 0.034(3)  & 0.705(1)  & 1.387(1)  & 0.364(1) & $-$0.115(9)   \\ \hline
\end{tabular} \end{center}
\end{table}

Because $p_c$ is very close to $p = 2$ for the XY model, one can expect that
the calculations based on higher order approximations will substantially
change the threshold dimension $d_c(p = 2)$, although for the Ising and
Heisenberg models, the changes of $d_c(p)$ should be small. This assumption
is supported by the calculation of critical indexes for three-dimensional
homogeneous models with $p = 1, 2, 3$ and the disordered Ising model. We
performed these calculations in the two-loop approximation with the use of
the Pade Borel summation technique (Table 4). The comparison of these
results with the corresponding indexes reported in \cite{16,17}, where the
all-time accurate calculations for threedimensional models were performed in
the six-loop approximation, shows that the difference in the values of
critical indexes does not exceed $0.02$.

The values of the threshold dimensions $d_c(p)$, which separate the domain
of critical behavior with RSB effects $d_c(p) < d < 4$ from the domain where
these effects are inessential, can be considered as threshold dimensions
that restrict the scope of the $\varepsilon$-expansion method as applied to
the three-vertex model of the weakly disordered system and the corresponding
results reported in \cite{4,5,6}. Our analysis also shows that the results
of application of the $\varepsilon$-expansion technique to multivertex
statistical models are unreliable independently of the approximation order.
This is explained by the competition between different types of fixed points
in the parametric space of multivertex models, which usually does not allow
one to pass to the limit as $\varepsilon \rightarrow 1$ without crossing the
marginal dimensions $3\leq d_c<4$ separating the stability domains of
different fixed points.

\begin{figure}[t]
\centering \includegraphics[height=10cm]{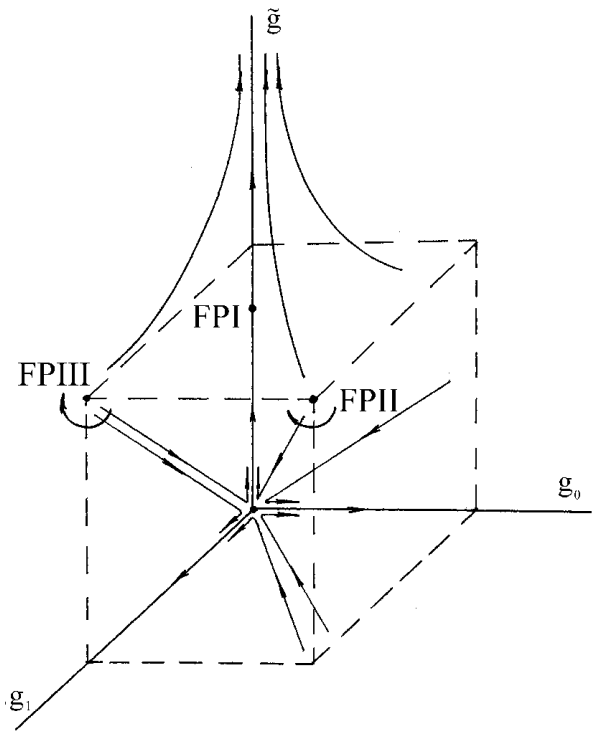}
\caption{The picture of renormalization group flows in the parametric space
($\tilde{g},g_0,g_1$) for the Ising model with the system dimension d =
3.99.}
\end{figure}

To reveal the character of the behavior of a disordered
system with RSB effects in the domain without
stable critical states, we analyzed the phase portrait of
the model based on the system of equations
\begin{equation}
r\frac{\partial g_i}{\partial r} = \beta_i( \tilde{g},g_0,g_1),
\end{equation}
which specifies phase trajectories in the space of vertices ($\tilde{g},g_0,g_1$). An
analysis shows (see Fig. 2) that, for the Ising model with $d_c = 3.986$ at $d \geq
3.986$, where none of the fixed points is stable, the strong coupling regime with the
renormalization group flows determined by ($\tilde{g},g_0,g_1$)$\rightarrow$($\infty
,0,0$) is realized if $\tilde{g} > \tilde{g}^*$. At the same time, at $\tilde{g} <
\tilde{g}^*$, the flows with ($\tilde{g},g_0,g_1$)$\rightarrow$($0,0,0$) are realized,
which asymptotically approach the Gauss fixed point $(0, 0, 0)$ and then also tend to
infinity along the axes $\tilde{g},g_0,g_1$. Such a behavior of the flows at $\tilde{g} <
\tilde{g}^*$ is caused by the closeness of the system dimension $d$ to four when the
effect of fluctuations is negligibly small and the Gauss fixed point becomes an
attractor.

\section{Conclusions}

The renormalization group analysis of weakly disordered
systems of an arbitrary dimension in the range
from three to four conducted in the two-loop approximation
showed that the critical behavior of threedimensional
systems is stable with respect to the effect
of the replica symmetry breaking. In systems with a
one-component order parameter, the critical behavior
determined by the structural disorder with a replica
symmetric fixed point is realized. The presence of weak
disorder does not affect the critical behavior of multicomponent
systems, although the proof of this fact for
systems with $p = 2$ requires calculations with higher
order approximations.

Effects of the replica symmetry breaking manifest
themselves only in disordered systems with the dimension
greater than three, and the threshold dimensions $d_c$
depend on the number of components of the order
parameter $p$ and the value of the parameter $x_0$. The predicted
picture of the influence of replica symmetry
breaking on the critical behavior of disordered systems
with a dimension $d > d_c$ qualitatively agrees with the
results reported in \cite{4,5,6}. The latter results were
obtained on the basis of the $\varepsilon$-expansion technique. For
systems with $p = 1$, RSB effects destroy the stable critical
behavior, and the strong coupling regime is realized;
for systems with $p = 2$ and $3$, a domain of nonuniversal
critical behavior occurs at $0 \leq x_0 \leq x_c(d)$. For $x_0$
outside this interval, the system exhibits no critical
behavior, as is the case at $p = 1$.

The values of threshold dimensions $d_c(p)$: $d_c(p = 1) =
3.986$, $d_c(p = 2) = 3.10$, and $d_c(p = 3) = 3.999$ which
separate the domain of critical behavior with RSB
effects $d_c(p) < d < 4$) from the domain where these
effects are insignificant, simultaneously specify the
lower bound of the domain where the results obtained
by $\varepsilon$-expansion are applicable to the description of the
model of weakly disordered systems with RSB effects
\cite{4,5,6}. It is noted that calculations carried out in higher
order approximations of the theory can significantly
change the threshold dimension dc for the XY model.
On the other hand, changes in $d_c(p)$ for the Ising and
Heisenberg models are expected to be small, which
leaves the scope of the results obtained by the $\varepsilon$-expansion
technique close to dimension four.

As the concentration of defects increases, one can expect a decrease of the threshold
values $d_c$ down to $d_c < 3$ beginning with a certain threshold concentration. In this
case, the influence of replica symmetry breaking effects can be significant. Due to
specific features of the manifestation of RSB effects, the concentration ns corresponding
to the spin percolation threshold can play the role of the threshold concentration of
defects for the Ising model, so that no stable critical behavior is observed at $n >
n_s$. For the XY and Heisenberg models, this role can be played by the concentration of
defects corresponding to the impurity percolation threshold $n_{imp} = 1 -  n_s$ with a
nonuniversal critical behavior for $n_{imp} < n < n_s$ and the absence of a stable
critical behavior at $n > n_s$.

\begin{center}
\bf Acknowledgmets
\end{center}
 This work was supported in part by the Ministry of Education of the
Russian Federation through Grants No. UR.01.01.052 and No. E02-3.2-196. \clearpage



\begin{thebibliography}{99}
\bibitem{1}
   S.F.~Edwards, P.W.~Anderson. J.Phys. {\bf F5}, 965 (1975).
\bibitem{2}
   J.~Emery. Phys. Rev. {\bf B11}, 239 (1975).
\bibitem{3}
   G.~Grinstein, A.~Luther. Phys. Rev. {\bf B13}, 1329 (1976).
\bibitem{4}
   Vik.S.Dotsenko, A.B.Harris, D.Sherrington, R.B.Stinchcombe.
   J.Phys.{\bf A28}, 3093 (1995).
\bibitem{5}
   Vik.S.~Dotsenko, D.E.~Feldman. J.Phys. {\bf A28}, 5183 (1995).
\bibitem{6}
   Vik.~S.~Dotsenko, Uspekhi Fizicheskikh Nauk, {\bf 165}, 481 (1995).
\bibitem{7}
   V.~V.~Prudnikov, A.~V.~Ivanov, A.~A.~Fedorenko,
      Sov. Phys. JETP Lett. {\bf 66} 835 (1997);
   V.~V.~Prudnikov, S.~V.~Belim, A.~V.~Ivanov, E.~V.~Osintsev,
      A.~A.~Fedorenko, Sov. Phys. JETP {\bf 87} 527 (1998);
   V.~V.~Prudnikov, P.~V.~Prudnikov, A.~A.~Fedorenko,
      Sov. Phys. JETP Lett. {\bf 68} 950 (1998);
      Phys.Rev. B {\bf 62}, 8777 (2000).
\bibitem{8}
   S.~A.~Antonenko, A.~I.~Sokolov, Phys.Rev. B {\bf 49}, 15901 (1994);
   A.I.~Sokolov, K.B.~Varnashev, A.I.~Mudrov. Int. J. Mod. Phys. B
      {\bf 12}, 1365 (1998);
   A.I.~Sokolov, K.B.~Varnashev, Phys.Rev. B {\bf 59}, 8363 (1999).
\bibitem{9}
V.~V.~Prudnikov, P.~V.~Prudnikov, A.~A.~Fedorenko,
      Sov. Phys. JETP Lett. {\bf 73},153 (2001).
\bibitem{10}
V.V.~Prudnikov, P.V.~Prudnikov, A.A.~Fedorenko. Phys.Rev. {\bf B63},
184201 (2001).
\bibitem{11}
A.~Pelissetto, E.~Vicari. Phys.Rev. {\bf B62}, 6393 (2000).
\bibitem{12}
V.~V.~Prudnikov, A.~N.~Vakilov, Sov. Phys. JETP {\bf 103}, 962 (1993).
\bibitem{13}
G.~Parisi. J.Phys. {\bf A13}, 1101 (1980);
   G.~Parisi. J.Phys. {\bf A13}, L115 (1980);
   G.~Parisi. J.Phys. {\bf A13}, 1887 (1980);
   M.~Mezard, G.~Parisi, M.~Virasoro. {\it Spin-Glass Theory and Beyond},
   Singapore: World Scientific, 1987;
   Vik.~S.~Dotsenko, Uspekhi Fizicheskikh Nauk {\bf 163}, 1 (1993).
\bibitem{14}
J.~Zinn-Justin. {\it Quantum Field Theory and Critical Phenomena}, Clarendon,
Oxford, 1996.
\bibitem{15}
M.~Dudka, Yu.~Holovatch, T.~Yavorskii. J.Phys.Stud. {\bf 5}, 233 (2001).
\bibitem{16}
J.C.~LeGuillou, J.~Zinn-Justin. Phys. Rev. {\bf B21}, 3976 (1980).
\bibitem{17}
A.~Pelissetto, E.~Vicari, e-print cond-mat/0002402.
\end{thebibliography}
\end{document}